%% file: arxiv-main.tex
\documentclass[a4paper]{article}
\pdfoutput=1

\usepackage[utf8]{inputenc}
\usepackage[nocompress]{cite}
\usepackage{fullpage}
\usepackage{hyperref}
\usepackage{amsmath, amssymb}
\usepackage{graphicx}
\usepackage{amsthm} 
\usepackage{xcolor}
\usepackage{cleveref}
\usepackage{fontawesome5}

\newtheorem{theorem}{Theorem}
\newtheorem{lemma}{Lemma}

\def\orcidsymbol{\includegraphics[height=9pt]{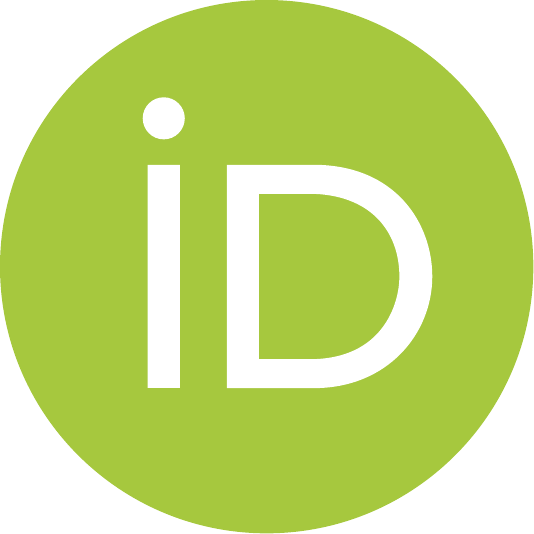}}
\def\mailsymbol{\textcolor{lightgray}{\fontsize{9}{12}\sffamily\bfseries \faIcon[regular]{envelope}}}

\date{}

\def\segments{\mathcal{L}}
\def\segaccess{\textsf{segment-access}}
\def\segsel{\textsf{segment-select}}
\def\segrank{\textsf{segment-rank}}

\def\slabrank{\textsf{slab-rank}}
\def\slabsel{\textsf{slab-select}}
\def\endrank{\textsf{endpoint-rank}}
\def\endsel{\textsf{endpoint-select}}

\begin{document}

\title{Succinct Data Structures for Segments}

\author{Philip Bille~\href{mailto:phbi@dtu.dk}{\mailsymbol}\href{https://orcid.org/0000-0002-1120-5154}{\orcidsymbol},
Inge Li G{\o}rtz~\href{mailto:inge@dtu.dk}{\mailsymbol}\href{https://orcid.org/0000-0002-8322-4952}{\orcidsymbol},
and Simon R. Tarnow~\href{mailto:sruta@dtu.dk}{\mailsymbol}\href{https://orcid.org/0009-0002-4293-6475}{\orcidsymbol}\\[0.5em]
{\small\begin{minipage}{\linewidth}\begin{center}
\begin{tabular}{c}
DTU Compute, Technical University of Denmark\\
Lyngby, Denmark\\
\end{tabular}
\end{center}\end{minipage}}
}

\maketitle

\input{./00-abstract.tex}
\input{./01-introduction.tex}
\input{./02-basics.tex}
\input{./03-segment-wavelet-tree.tex}

\input{./04-slab.tex}
\input{./05-lower-bounds.tex}

%% Bibliography
%%
\bibliography{bibliography.bib}
\end{document}

%% file: 00-abstract.tex
\begin{abstract}
We consider succinct data structures for representing a set of $n$ horizontal line segments in the plane given in rank space to support \emph{segment access}, \emph{segment selection}, and \emph{segment rank} queries.  A segment access query finds the segment $(x_1, x_2, y)$ given its $y$-coordinate ($y$-coordinates of the segments are distinct), a segment selection query finds the $j$th smallest segment (the segment with the $j$th smallest $y$-coordinate) among the segments crossing the vertical line for a given $x$-coordinate, and a segment rank query finds the number of segments crossing the vertical line through $x$-coordinate $i$ with $y$-coordinate at most $y$, for a given $x$ and $y$. This problem is a central component in compressed data structures for persistent strings supporting random access. 

Our main result is data structure using $2n\lg{n} + O(n\lg{n}/\lg{\lg{n}})$ bits of space and $O(\lg{n}/\lg{\lg{n}})$ query time for all operations. We show that this space bound is optimal up to lower-order terms. We will also show that the query time for segment rank is optimal. The query time for segment selection is also optimal by a previous bound. 

To obtain our results, we present a novel segment wavelet tree data structure of independent interest. This structure is inspired by and extends the classic wavelet tree for sequences. This leads to a simple, succinct solution with $O(\log n)$ query times. We then extend this solution to obtain optimal query time. Our space lower bound follows from a simple counting argument, and our lower bound for segment rank is obtained by a reduction from 2-dimensional counting.   
\end{abstract}

%% file: 01-introduction.tex
\section{Introduction}
Let $\segments$ be a set of $n$ horizontal line segments in \emph{rank space},
that is, the line segments are in the plane $[1,2n]\times [1,n]$ such that there is exactly one endpoint on each $x$-coordinate and one segment on each $y$-coordinate.
The \emph{segment representation problem} is to preprocess $\segments$ to support the operations:
\begin{itemize}
    \item $\segaccess(y)$: return the segment with $y$-coordinate $y$.
    \item $\segsel(i,j)$: return the $y$-coordinate of the $j$th smallest segment (the segment with the $j$th smallest $y$-coordinate) among the segments crossing the vertical line through $x$-coordinate $i$.
    \item $\segrank(i,y)$: return the number of segments crossing the vertical line through $x$-coordinate $i$ with $y$-coordinate at most $y$.
\end{itemize}
Here, we consider a segment $(x_l, x_r, y)$ to be crossing the vertical line through $x$-coordinate $i$ iff $x_l\leq i < x_r$.
The segment representation problem in which the endpoints are real numbers can be reduced to the rank space variant using standard techniques \cite{gabow1984scaling}.
Bille and Gørtz~\cite{DBLP:journals/mst/BilleG23} considered representing segments to support segment selection queries in connection with compressed data structures for persistent strings. Here, the problem is supporting random access on a set of strings represented by a version tree, where each edge represents a replace, insert, or delete operation on the string represented by the parent node. They showed that this problem can be reduced to answering segment selection queries on horizontal line segments. They gave a data structure supporting segment selection queries using $O(n\lg{n})$ bits of space and $O(\lg{n}/\lg{\lg{n}})$ query time\footnote{We denote $\lg{n}=\log_2{n}$.}. Furthermore, they showed that $\Omega(\lg{n}/\lg{\lg{n}})$ time is required to answer segment selection queries for any static data structure using $n\lg^{O(1)}{n}$ bits of space.

\subsection{Results}
This paper considers succinct data structures for the segment representation problem. We show the following main result on a standard unit cost word RAM model with logarithmic word size. 
\begin{theorem}\label{theorem:segment_select_fusion}
  Given a set of $n$ horizontal line segments, we can solve the segment representation problem using $2n\lg{n} + O(n\lg{n}/\lg{\lg{n}})$ bits of space and $O(\lg{n}/\lg{\lg{n}})$ time for all queries.
\end{theorem}
Compared to previous results of Bille and Gørtz~\cite{DBLP:journals/mst/BilleG23}, \cref{theorem:segment_select_fusion} improves the space bounds from $O(n\lg{n})$ to $2n\lg{n} + O(n\lg{n}/\lg{\lg{n}})$. At the same time, we obtain the optimal $O(\lg{n}/\lg{\lg{n}})$ query time for segment selection and implement the segment rank query in the same time. Furthermore, 
we show that the space bound of \cref{theorem:segment_select_fusion} is optimal up to lower order terms. 
\begin{theorem}\label{theorem:segment:succinct}
Any data structure representing $n$ horizontal line segments requires at least $2n\lg{n}-O(n)$ bits.
\end{theorem}
Finally, we show that the query time for segment rank is also optimal.
\begin{theorem}\label{tm:segrank_lowerbound}
    Any static data structure on $n$ horizontal line segments that uses $n\lg^{O(1)}{n}$ bits of space needs $\Omega(\frac{\lg{n}}{\lg{\lg{n}}})$ time to support segment rank queries.
\end{theorem}

\subsection{Techniques}
We obtain our results by first considering a novel and simple structure, the \emph{segment wavelet tree}, which may be of independent interest. 
The segment wavelet tree is inspired by the classical wavelet tree structure of Grossi et al.~\cite{DBLP:conf/soda/GrossiGV03},
and builds on the observation that the number of segments crossing a vertical line is the difference between the number of left and right endpoints occurring before said line.
In the same manner, as the wavelet tree recursively splits the alphabet in its lower and upper half, the segment wavelet tree recursively splits the segments in the lower and upper half of the plane.
Because of this, the segment wavelet tree can be used to search for the $j$th segment crossing a vertical line.
The segment wavelet tree, however, only achieves $O(\lg{n})$ query time since it only splits the plane into $2$ horizontal bands.
In order to speed up the query to $O(\lg{n}/\lg{\lg{n}})$ we generalize the segment wavelet tree to the \emph{$\Delta$-ary segment wavelet tree}, which splits the plane into $\Delta$ horizontal bands called \emph{slabs}, leading to \cref{theorem:segment_select_fusion}.
Next, we prove the information-theoretical lower bound of representing horizontal line segments by showing that in rank space, we need at least $2n\lg{n} - O(n)$ bits of space to represent $n$ horizontal line segments.
Finally, we prove a matching lower bound for the segment rank query on horizontal line segments by showing that any static solution using $n\lg^{O(1)}{n}$ bits of space needs $\Omega(\lg{n}/\lg{\lg{n}})$ query time.
To do so, we show a reduction from \emph{2-dimensional dominance counting}~\cite{DBLP:conf/stoc/Patrascu07}.

\subsection{Related Work}
\subparagraph*{Queries on Intervals}
There exists a related problem called the \emph{stabbing-semigroup problem}.
The stabbing-semigroup problem is to preprocess a set of $n$ intervals where each interval has an associated weight, such that given an integer $x$, we can compute the sum of weights of the intervals containing $x$.
Agarwal et al.~\cite{DBLP:journals/siamcomp/AgarwalAKMTY12} showed how to solve the stabbing-semigroup problem in $O(n\lg{n})$ bits of space and $O(\lg{n})$ query time.
They also showed how to make the structure dynamic, allowing adding and removing intervals in $O(\lg{n})$ time and how it can be adapted to work in external memory.
Another related query on intervals is the stabbing-max, which is the problem of finding the interval of maximum weight containing $x$.
Nekrich~\cite{DBLP:journals/corr/abs-1109-3890} described a dynamic data structure that answers one-dimensional stabbing-max queries in optimal time $O(\lg{n}/\lg{\lg{n}})$ using $O(n\lg{n})$ bits and allows insertions and deletions of intervals in $O(\lg{n})$ time.
To the best of our knowledge, neither of these problems has been considered in a succinct setting.

\subparagraph*{Succinct Data Structures}
Many geometric queries on two-dimensional points have been considered in a succinct setting, including orthogonal range reporting and counting~\cite{DBLP:journals/siamcomp/Chazelle88,DBLP:conf/isaac/JaJaMS04,DBLP:journals/tcs/MakinenN07}, point location~\cite{DBLP:journals/talg/BoseCHMM12,DBLP:conf/isaac/HeNZ12} and data-analysis queries~\cite{DBLP:journals/tcs/NavarroNR13} (see also the survey by He~\cite{DBLP:conf/birthday/He13}).
We extend this line of research by considering horizontal line segments in a succinct setting.

\subparagraph*{2-Dimensional Dominance Counting}
The \emph{2-dimensional dominance counting problem} is to preprocess $n$ points, such that given a point $(x,y)$, we can compute the number of points $(x', y')$ where $x' \leq x$ and $y'\leq y$.
The $\segrank(i,y)$ can also be viewed as two $2$-dimensional dominance counting queries by observing that the number of segments crossing the vertical line through $x$-coordinate $i$ with $y$-coordinate at most $y$ is the difference in the number of left and right endpoints dominating $(i,y)$.
Bose et al.~\cite{DBLP:conf/wads/BoseHMM09} showed that we can answer 2-dimensional dominance counting queries in $O(\lg{n}/\lg{\lg{n}})$ time using $n\lg{n} + o(n\lg{n})$ space, when the points are in rank space.
Thus, we can achieve the same results for segment rank queries as in \cref{theorem:segment_select_fusion} using two 2-dimensional dominance counting structures.
However, this leaves out how to answer segment access and selection queries.

\subparagraph*{Range Selection}
The \emph{range selection problem} is to preprocess an array $A$ of $n$ unique integers, such that given a query $(i,j,k)$, one can report the $k$th smallest integer in the subarray $A[i], A[i+1],\ldots, A[j]$.
A slight variation of the range selection problem is the \emph{prefix selection problem}, which fixates $i=1$ in the query.
Due to Jørgensen and Larsen~\cite{DBLP:conf/soda/JorgensenL11}, the prefix selection problem can be solved in $O(n\lg{n})$ bits and $O(\lg{n}/\lg{\lg{n}})$ query time with matching lower bounds.
Bille and Gørtz~\cite{DBLP:journals/mst/BilleG23} showed the similarity between prefix selection and segment selection by proving that $\Omega(\lg{n}/\lg{\lg{n}})$ time is required to answer segment selection queries for any static data structure using $n\lg^{O(1)}{n}$ space, by reduction from prefix selection.

\subsection{Outline}
We present the required preliminaries for our solution in \cref{section:preliminaries}, and then we describe a simple structure in \cref{section:segment-wavelet-tree} and how we can answer segment rank and select queries with this structure.
We then show our structure in \cref{section:slab} and how we answer rank and select queries succinctly in $O(\lg{n}/\lg{\lg{n}})$ time leading to \cref{theorem:segment_select_fusion}.
Finally, we prove the space lower bounds for representing segments in rank space leading to \cref{theorem:segment:succinct} and prove the lower bounds for segment rank queries leading to \cref{tm:segrank_lowerbound} in \cref{section:lowerbounds}.

%% file: 02-basics.tex
\section{Preliminaries}\label{section:preliminaries}
For a given problem $P$ with $|\mathcal{U}|$ instances, we need $\lceil\lg{|\mathcal{U}|}\rceil$ bits to distinguish between each instance in the worst case.
We say a data structure representing $P$ is \emph{compact} if it uses $O(\lceil\lg{|\mathcal{U}|}\rceil)$ bits of space and is \emph{succinct} if it uses $\lceil\lg{|\mathcal{U}|}\rceil+o(\lg{|\mathcal{U}|})$ bits of space \cite{jacobson1988succinct}.

\subsection{Succinct Representations of Strings}
Let $S[1,n]$ be a string of $n$ characters from an alphabet $\Sigma=\{1,2,\ldots, \sigma\}$ and define the following operations.
\begin{itemize}
\item $\textsf{access}(S,i):$ return $S[i]$.
\item $\textsf{rank}(S, \alpha, i):$ return the number of occurrences of character $\alpha$ in $S[1,i]$.
\item $\textsf{select}(S, \alpha, i):$ return the position in $S$ of the $i$th occurrence of character $\alpha$. 
\end{itemize}

For binary strings, we use the following well-known result:
\begin{lemma}[\cite{DBLP:conf/focs/Jacobson89}]\label{lemma:bitvector}
We can represent a bit string of length $n$ using $n + o(n)$ bits and support  \textsf{access}, \textsf{rank}, and \textsf{select} queries in constant time.
\end{lemma}

\subparagraph*{Wavelet Tree}
A \emph{wavelet tree} \cite{DBLP:conf/soda/GrossiGV03} for a string $S$ is a complete balanced binary tree $T$ with $\sigma$ leaves. 
Each node $v$ in $T$ represents the characters in $\Sigma(v)=[a,b]\subseteq \Sigma$.
The root represents the full alphabet $\Sigma$ and for any non-leaf node $v$ with  alphabet $\Sigma(v)=[a,b]$ the left child $v_0$ 
represents the lower half of $\Sigma(v)$, i.e.,  $\Sigma(v_0)=[a,a + \lfloor (b-a)/2\rfloor]$, and the right child $v_1$  the upper half of $\Sigma(v)$, i.e., $\Sigma(v_1)=[a+\lfloor (b-a)/2\rfloor + 1, b]$.
Furthermore, for any leaf node $v$, we have $\Sigma(v)=[c,c]$.
For a node $v$ let $S(v)$ be $S$ restricted to the characters $\Sigma(v)$.
Each internal node $v$ with left child $v_0$ and right child $v_1$ stores a bitstring $B(v)$ such that $B(v)[i] = 0$ if $S(v)[i] \in \Sigma(v_0)$ and $B(v)[i] = 1$ otherwise ($S(v)[i] \in \Sigma(v_1)$).

With the bit string $B(v)$ we can track which child of $v$ will contain each element $S(v)[i]$, by observing that child $v_b$, where $b\in[0,1]$, will contain element $S(v)[i]$ iff $B(v)[i] = b$.
If $B(v)[i]=b$ then $S(v)[i]$ will be stored at $S(v_b)[j]$ where index $j$ is the number of occurrences of $b$ in $B(v)[1,i]$ since each occurrence of $b$ is an element in $S(v)$ that would also be in $S(v_b)$ and would be in the same order as they appear in $S(v)$. 
This is exactly the \textsf{rank} operation on bit strings, thus $S(v)[i]=S(v_b)[\textsf{rank}(B(v), b, i)]$.
With this property, we can navigate downwards in the wavelet tree.
We can also use the bit string $B(v)$ together with \textsf{select} to navigate upwards in the wavelet tree. Let $v_b$ be a child of $v$. Then the index $i$ that the symbol $S(v_b)[j]$ occurs in $S(v)$ is the index of the $j$th $b$ in $B(v)$ which is $i=\textsf{select}(B(v), b, j)$.
The following lemma captures these properties:
\begin{lemma}\label{lemma:wavelet_mapping}
Let $v$ be a non-leaf node in a wavelet tree. Given an index $i \in [1,|B(v)|]$ then $j=\textsf{rank}(B(v), b, i)$ is the greatest index in the bit string of the child $v_b$ such that $\textsf{select}(B(v), b, j) \leq i$.
In particular, if $B(v)[i]=b$ then $\textsf{select}(B(v),b,j)=i$.
\end{lemma}
This follows from the duality of \textsf{rank} and \textsf{select} and the observations made above.
Using \cref{lemma:bitvector} to represent the bit strings of each node one can implement a wavelet tree for a sequence $S[1,n]$ over the alphabet $\Sigma$ using $n\lceil\lg{\sigma}\rceil + o(n\lg{\sigma})$ bits of space and $O(\lg{\sigma})$ query time for \textsf{rank}, \textsf{select}, and \textsf{access} \cite{DBLP:journals/tcs/MakinenN07}.
A more recent result shows the following:
\begin{lemma}[\cite{DBLP:conf/esa/GolynskiGGRR07}]\label{lemma:wavelet:space}
  The wavelet tree for a sequence $S[1,n]$ over the alphabet $\Sigma=\{1,2,\ldots, \sigma\}$ uses $n\lceil\lg{\sigma}\rceil + o(n)$ bits of space and $O(\lg{\sigma})$ query time for \textsf{rank}, \textsf{select}, and \textsf{access}.
\end{lemma}

%% file: 03-segment-wavelet-tree.tex
\section{Segment Wavelet Tree}\label{section:segment-wavelet-tree}
Here, we present a simple succinct solution to the segment representation problem with the following bounds.
\begin{theorem}\label{theorem:segment_select}
Given a set of $n$ horizontal line segments in the plane $[1,2n]\times[1,n]$, we can solve the segment representation problem succinctly in $2n\lg{n} + O(n)$ bits and $O(\lg{n})$ time for all queries.
\end{theorem}
We will begin by describing the content of our data structure and then show how we can answer queries using this structure.

\subsection{Data Structure}\label{section:datastructure}
Let $\segments$ be a set of $n$ horizontal line segments in rank space. We define the segment wavelet tree as a recursive decomposition of the line segments in $\segments$ similar to the wavelet tree. 
For simplicity, we assume that $n$ is a power of $2$.
Define $\segments[a,b] \subseteq \segments$ to be the segments with $y$-coordinate in $[a,b]$. The \emph{segment wavelet tree} $T$ for $
\segments$ is a complete balanced binary tree on $n$ leaves. Each node $v$ represents a set of segments $\segments(v) = \segments[a,b]$. The root $r$ represents $\segments(r) = \segments[1,n] = \segments$. Let $v$ be an internal node representing the segments $\segments(v) = \segments[a,b]$, and let $v_0$ and $v_1$ be the left and right child of $v$, respectively. Then $v_0$ represents the segments $\segments(v_0) = \segments[a,\lfloor (b-a)/2\rfloor]$ and $v_1$ represents the segments 
$\segments(v_1) = \segments[\lfloor (b-a)/2\rfloor+1, b]$. A leaf node $v$ represents a single segment $\segments(v) = \segments[a,a]$. 

\begin{figure}[!ht]
    \begin{center}
    \includegraphics[scale=0.8]{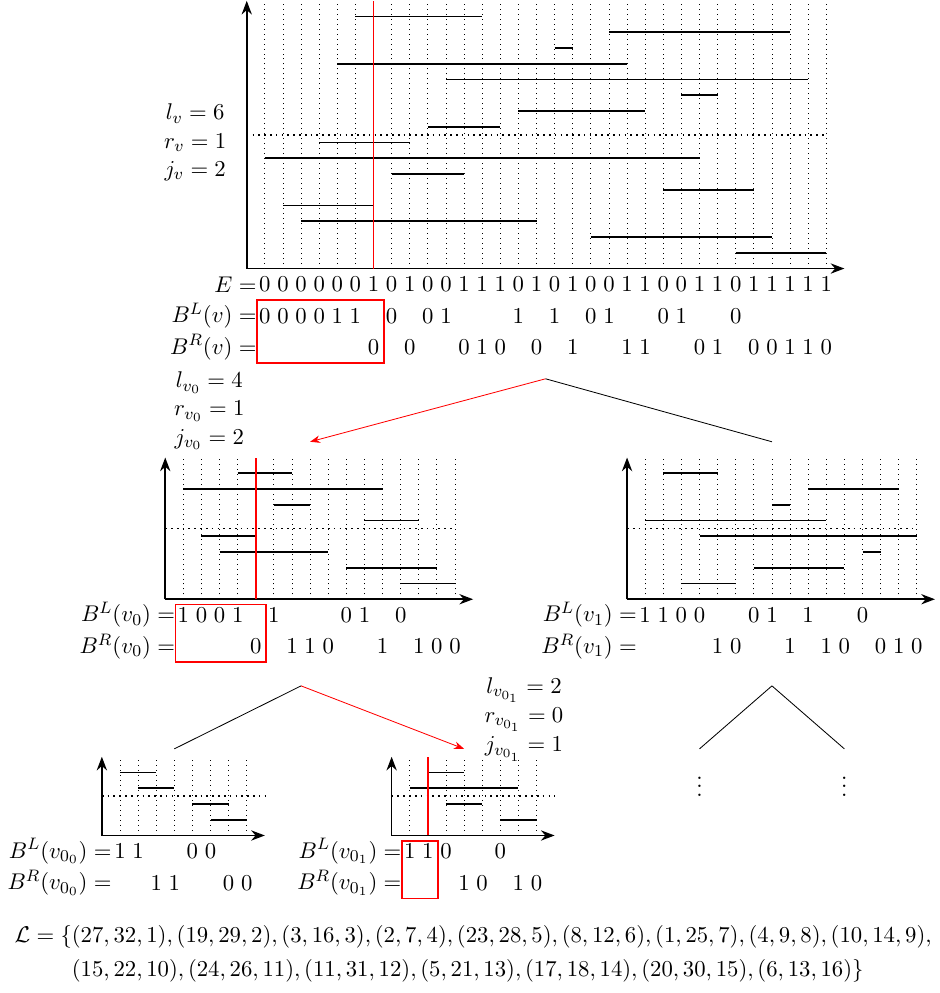}
    \caption{The top 3 levels of the segment wavelet tree of the segments $\segments$ and the computed local variables for the query $\segsel(7,2)$, 
    where $v$ is root of the segment wavelet tree.
    For some of the nodes, the corresponding subproblem is visualized as a 2D plane, where empty columns have been removed.
    The bitvectors $B^L$ and $B^R$ of each node is horizontally spaced such that each bit vertically aligns with the endpoint it represents.
    The visited nodes in the query $\segsel(7,2)$ are marked with a red arrow together with the local variables.
    Furthermore, in the 2D plane of the visited nodes, the vertical line with $x$-coordinate $7$ is highlighted,
    and the prefix of the bitvectors $B^L$ and $B^R$ that correspond to the endpoints with $x$-coordinate at most $7$ are also highlighted.
    }
    \label{fig:example_query}
    \end{center}
\end{figure}

We store the segment wavelet tree succinctly as follows. Each internal node $v$ stores two bitstrings $B^L(v)$ and $B^R(v)$ of length $|\segments(v)|$. 
\begin{align*}
    B^L(v)[i] &= 
    \begin{cases}
        0 & \text{if the $i$th segment in $\segments(v)$ ordered by left endpoint is in $\segments(v_0)$}\\
        1 & \text{otherwise}
    \end{cases} \\
     B^R(v)[i] &= 
    \begin{cases}
        0 & \text{if the $i$th segment in $\segments(v)$ ordered by right endpoint is in $\segments(v_0)$}\\
        1 & \text{otherwise}
    \end{cases} 
\end{align*}
Furthermore, we store a bitstring $E[1,2n]$ where 
\begin{align*}
    E[i] &= 
    \begin{cases}
        0 & \text{there is a left endpoint with $x$-coordinate $i$ in~$\segments$}\\
        1 & \text{otherwise}
    \end{cases}
\end{align*}
See \cref{fig:example_query} for an example. 

Alternatively, one can also view the segment wavelet tree as two superimposed wavelet trees of the strings $Y^L[1,n]$ and $Y^R[1,n]$, 
where $Y^L[1,n]$ and $Y^R[1,n]$  are the strings of $y$-coordinates of the left and right endpoints, respectively, ordered by increasing $x$-coordinate.

To achieve the desired space bounds, we store $B^L$ and $B^R$ as their superimposed wavelet trees $Y^L$ and $Y^R$ according to \cref{lemma:wavelet:space} and the bitstring $E$ according to \cref{lemma:bitvector}.

\subparagraph*{Analysis} The total length of $Y^L$ and $Y^R$ is $2n$ and the length of $E$ is $2n$. By \cref{lemma:bitvector,lemma:wavelet:space} the total space is $2n\lceil\lg{n}\rceil + o(n) + 2n + o(n) = 2n\lg{n}+ O(n)$ bits.

\subsection{Segment Access Queries}
We now show how to answer
$\segaccess$ queries in $O(\lg{n})$ time.
To answer a $\segaccess(y)$ query, we do a bottom-up traversal of the segment wavelet tree $T$ starting at the $y$th leaf in the left-to-right order. Let $s = (x_l, x_r, y)$ denote the segment we are searching for. At each node $v$ in the traversal we maintain the following \emph{local variables}:
\begin{align*}
    l_v &= \text{the number of segments in $\segments(v)$ whose left endpoint has $x$-coordinate at most $x_l$.}\\
    r_v &= \text{the number of segments in $\segments(v)$ whose  right endpoint has $x$-coordinate at most $x_r$.}
\end{align*}
We perform the traversal as follows.
Initially, $v$ is the $y$th leaf and $l_v = r_v = 1$.
Consider a non-root node $v$ with parent $v_p$ and let $b = 0$ if $v$ is the left child of $v_p$ and $b=1$ otherwise.
We compute $l_{v_p} = \mathsf{select}(B^L(v_p), b, l_v)$ and $r_{v_p} = \mathsf{select}(B^R(v_p), b, r_v)$.
When we reach the root $u$ we compute $x_l = \textsf{select}(E, 0, l_u)$ and $x_r= \textsf{select}(E, 1, r_u)$. Finally, we return $(x_l, x_r, y)$.

\subparagraph*{Analysis} We use constant time at each node, and we visit one node at each level of the segment wavelet tree. Since the height of $T$ is $O(\lg{n})$, the total time is $O(\lg{n})$.

\subparagraph*{Correctness}
Let $s=(x_l, x_r, y)$ be the segment with $y$-coordinate $y$.
We show inductively that $l_v$ and $r_v$ are computed correctly for each $v$ on the path in the bottom-up traversal.
When $v$ is the $y$th leaf we have $\segments(v)=\{s\}$ and thus $l_v=1$ and $r_v=1$.
Consider an internal non-root node $v$.
Assume $s\in\segments(v)$ and $l_v$ and $r_v$ are correct.
Let $v_p$ be the parent of $v$ and $b=0$ if $v$ is the left child of $v_p$ and $b=1$ otherwise.
The bitstrings $B^L(v_p)$ and $B^R(v_p)$ are $b$ in every index corresponding to a segment in $\segments(v)$.
Thus $l_{v_p} = \mathsf{select}(B^L(v_p), b, l_v)$ is the number of segments in $\segments(v_p)$ whose left endpoint has $x$-coordinate at most $x_l$, and $r_{v_p} = \mathsf{select}(B^R(v_p), b, r_v)$ is the number of segments in $\segments(v_p)$ whose right endpoint has $x$-coordinate at most $x_r$.
Let $u$ be the root.
Since $E$ are $0$ and $1$ in every index corresponding to a left and right segment endpoint, respectively, we have $x_l = \textsf{select}(E, 0, l_u)$ and $x_r= \textsf{select}(E, 1, r_u)$.

\subsection{Segment Select Queries}
\label{sec:segment_wavelet:select}
We now show how to answer 
$\segsel$ queries in $O(\lg{n})$ time.
To answer a $\segsel(i,j)$ query, we do a top-down traversal in the segment wavelet tree $T$  starting at the root and ending in the leaf containing the $j$th crossing segment of the vertical line at time $i$.

At each node $v$ with $\segments(v) = \segments[a,b]$ in the traversal we maintain the following \emph{local variables}:
\begin{align*}
    l_v =& \text{the number of segments in $\segments(v)$ whose  left endpoint has $x$-coordinate at most $i$.}\\
    r_v =& \text{the number of segments in $\segments(v)$ whose  right endpoint has $x$-coordinate at most $i$.}\\
    \bar{j}_v =& \text{the number of segments in $\segments[1, a-1]$ crossing the vertical line at $i$.}\\
    j_v =& \text{$j - \bar{j}_v$, such that the $j_v$th segment in $\segments(v)$ crossing the vertical line at $i$}\\
    &\text{is the $j$th segment in $\segments$ crossing the vertical line at $i$.} 
\end{align*}
We perform the traversal as follows. 
Initially, $v$ is the root and we have $r_{v}=\textsf{rank}(E, 1, i)$, $l_{v}=i-r_{v}, \bar{j}_{v}=0$ and $j_{v}=j$.
Consider an internal node $v$ with children $v_0$ and $v_1$ and suppose we have computed $l_v$, $r_v, \bar{j}_{v}$, and $j_v$. 
We first compute the number of segments $k$ in $\segments(v_0)$ crossing the vertical line at $i$ as
\[
k = \textsf{rank}(B^L(v), 0, l_v) - \textsf{rank}(B^R(v), 0, r_v)\;.
\]
That is, $k$ is computed as the number of left endpoints of segments in $\segments(v_0)$ with $x$-coordinate at most $i$ subtracted by the number of right endpoints in $\segments(v_0)$ with $x$-coordinate at most $i$.

We continue the traversal in child $v_b$, where $b = 0$ if $j_v \leq k$ and $b=1$ otherwise. 

We then compute the local variables for the child $v_b$ as
    \begin{align*}
        l_{v_b} &= \textsf{rank}(B^L(v), b, l_v)\\
        r_{v_b} &= \textsf{rank}(B^R(v), b, r_v)\\
        \bar{j}_{v_b} &= \begin{cases}
            \bar{j}_v & \text{if }b=0 \\
            \bar{j}_v + k &\text{otherwise}
        \end{cases}\\
        j_{v_b} &= j - \bar{j}_{v_b}
    \end{align*}
When $v$ is a leaf and $\segments(v)=\segments[a,a]$ we return $a$. 

\subparagraph*{Analysis} We use constant time at each node and visit one node at each level of the segment wavelet tree. Since the height of $T$ is $O(\lg{n})$, the total time is $O(\lg{n})$.

\subparagraph*{Correctness} 
Let $s$ be the $j$th smallest segment crossing the vertical line at $i$. 
We show inductively, that $s\in \segments(v)$ and $l_v$, $r_v$, $\bar{j}_v$, and $j_v$ are computed correctly for each $v$ on the path in the top-down traversal.

When $v$ is the root we have $s\in \segments(v) = \segments$ and $\bar{j}_v=0$ and $j_v=j$. By definition, we have $r_v= \textsf{rank}(E,1,i)$ is the number of segments in $\segments$ whose right endpoint has $x$-coordinate at most $i$. Then, $l_v = i - r_v = \textsf{rank}(E,0,i)$ is the number of segments in $\segments$ whose left endpoint has $x$-coordinate at most $i$. 

Consider an internal non-root node $v$. Assume $s\in \segments(v)$ and that $l_v$, $r_v$, $\bar{j}_v$, and $j_v$ are correct. Let $v_b$ be the child of $v$ computed by the algorithm. The bitstrings $B^L(v)$ and $B^R(v)$ are $0$ in every index corresponding to a segment in $\segments(v_0)$.
Thus $k= \textsf{rank}(B^L(v), 0, l_v) - \textsf{rank}(B^R(v), 0, r_v)$ is the number of segments crossing the vertical line at $i$ in $\segments(v_0)$. The set $\segments(v_0)$ contains the lower half of the segments in $\segments(v)$. Thus if $j_v \leq k$ then $s\in \segments(v_0)$, $\bar{j}_{v_0}=\bar{j}_v$ and $j_{v_0} = j_v = j - \bar{j}_{v_0}$. Otherwise,  $s\in \segments(v_1)$, $\bar{j}_{v_1} = \bar{j}_v + k$ and $j_{v_1} = j_v - k = j - \bar{j}_{v_1}$. It follows that we continue the traversal in the correct child $v_b$. By definition of $B^L(v)$ and $B^R(v)$ it follows that  $l_{v_b}$, and $r_{v_b}$ are computed correctly. 

\subsection{Segment Rank Queries}
We now show how to answer $\segrank$ queries in $O(\lg{n})$ time.
To answer a $\segrank(i, y)$ we perform a top-down traversal in the segment wavelet tree $T$ starting at the root and ending in the leaf $v$ such that $\segments(v)=\segments[y,y]$.
At each node $v$ in the traversal we compute the local variables $l_v$, $r_v$ and $\bar{j}_v$ in the same manner as in \cref{sec:segment_wavelet:select}.
We perform the traversal as follows.
Consider an internal node $v$ with $\segments(v)=\segments[a,b]$ and children $v_0$ and $v_1$,
we continue the traversal in child $v_b$, where $b=0$ if $y \leq \lfloor(b-a)/2\rfloor$ and $b=1$ otherwise.
When $v$ is a leaf we return $\bar{j_v} + (l_v - r_v)$.

\subparagraph*{Analysis} We use constant time at each node, and we visit one node at each level of the segment wavelet tree. Since the height of $T$ is $O(\lg{n})$, the total time is $O(\lg{n})$.

\subparagraph*{Correctness}
The correctness follows immediately from the correctness argument in \cref{sec:segment_wavelet:select} and the definition of $l_v$, $r_v$ and $\bar{j_v}$.

\medskip
In summary, we have shown \cref{theorem:segment_select}.

%% file: 04-slab.tex
\section{Generalized Segment Wavelet Tree}\label{section:slab}
Here, we generalize the solution in \cref{section:segment-wavelet-tree} to a tree of out-degree $\Delta=\lceil\lg^\epsilon{n}\rceil$, where $0 < \epsilon < 1$.
To increase the out-degree to $\Delta$, we first consider the required data structure to partition the segments into $\Delta$ horizontal bands called \emph{slabs}.
Afterward, we will describe the content of our data structure and show how we can answer queries using this structure.

\subsection{Succinct Slab Representation}
Let $\segments$ be a set of $n$ horizontal line segments partitioned into $\Delta=\lceil\lg^\epsilon{n}\rceil$ slabs of approximately equal size, where $0 < \epsilon < 1$.
The \emph{slab representation problem} is to preprocess $\segments$ to support the operations:
\begin{itemize}
    \item $\slabsel(v,i,j)$: return the slab $k$ containing the $j$th segment in $\segments(v)$ according to increasing $y$-coordinate among the segments crossing the vertical line through $x$-coordinate $i$.
    \item $\slabrank(v,i,j)$: return the number of segments in $\segments(v)$ crossing the vertical line through $x$-coordinate $i$ in slabs $[1,j]$.
    \item $\endsel(v, k, i)$: return the $x$-coordinate of the $i$th endpoint in $\segments(v)$ in the $k$th slab according to increasing $x$-coordinate among the segments.
    \item $\endrank(v, k, i)$: return the number of endpoints in $\segments(v)$ in the $k$th slab who has a $x$-coordinate of at most $i$.
\end{itemize}

\begin{lemma}[\cite{DBLP:journals/mst/BilleG23}]\label{lemma:narrowgrid}
    Given a set of $n$ horizontal line segments, partitioned into $\Delta=\lceil{\lg^\epsilon{n}}\rceil$ horizontal slabs for $0< \epsilon < 1$, we can solve the slab representation problem in $2n\lg{\lg^\epsilon{n}} +O(n)$ bits of space and $O(1)$ time for all queries.
\end{lemma}
\begin{proof}
In \cite{DBLP:journals/mst/BilleG23} they show that $\slabsel$ and $\slabrank$ can be done in $O(n\lg{\lg^\epsilon{n}})$ space and constant query time using $O(n)$ preprocessing time.
We show how to modify this result and the analysis to achieve $2n\lg{\lg^\epsilon{n}} +O(n)$ bits of space while maintaining constant query time.
They partition the sequence of segment endpoints into blocks, which are further partitioned into cells. 
Their data structure consists of the following components.
\begin{enumerate}
    \item A predecessor data structure for each block.
    \item The first column of each cell.
    \item The sequence of slab indices each endpoint belongs to, ordered by increasing $x$-coordinate.
    \item A global table for tabulating queries inside the cells.
\end{enumerate}
We modify the block width to be ${\lceil\lg^\lambda{n}\rceil\lceil\lg{n}\rceil}$
instead of ${\lceil\lg^\epsilon{n}\rceil\lceil\lg{n}\rceil}$, for another parameter $\epsilon < \lambda < 1$.
This also sets the cell width to $\lceil\lg^\lambda{n}\rceil$.
Plugging into their analysis, this implies the following space bounds.
\begin{enumerate}
    \item  The predecessor structure uses $O(\lg^\epsilon{n}\lg{n})$ space for each block. The number of blocks is $\frac{2n}{\lceil\lg^\lambda{n}\rceil\lceil\lg{n}\rceil}$, hence the space required is $O(2n\frac{\lg^\epsilon{n}}{\lg^\lambda{n}})=o(n)$.
    
    \item Each entry of the first column of a cell can be encoded in $O(\lg{\lg^\epsilon{n}})$.
    The combined height of all cells is $2n\frac{\lceil{\lg^\epsilon{n}}\rceil}{\lceil\lg^\lambda{n}\rceil}$ and thus the combined space is $2n\frac{\lceil{\lg^\epsilon{n}}\rceil}{\lceil\lg^\lambda{n}\rceil}\cdot O(\lg{\lg^\epsilon{n}})=o(n)$.
    
    \item The sequence contains $2n$ endpoints, and each endpoint can be encoded in $\lg{\lg^\epsilon{n}} + O(1)$ bits, thus the entire sequence uses $2n\lg{\lg^\epsilon{n}} + O(n)$ space.
    
    \item The global table has size $2^{O((\lg^\lambda{n} + \lg^\epsilon{n})\cdot \lg{\lg{n}})} \cdot O(\lg{\lg^\epsilon{n}}) = 2^{O((\lg^\lambda{n} + \lg^\epsilon{n})\cdot \lg{\lg{n}})}=o(n)$.
\end{enumerate}
In total, the data structure uses $2n\lg{\lg^\epsilon{n}} + O(n)$ space and the same $O(n)$ preprocessing time.
Using standard techniques~\cite{DBLP:journals/talg/FerraginaMMN07}, we can also support $\textsf{select}$ and $\textsf{rank}$ on the sequence of endpoints in constant time using $O(n)$ extra space.
Since the sequence of endpoints is ordered by increasing $x$-coordinate $\textsf{select}$ and $\textsf{rank}$ are equivalent with $\endsel$ and $\endrank$, respectively.
\end{proof}

\subsection{Data Structure}\label{sec:multi:datastructure}
Let $\segments$ be a set of $n$ horizontal line segments in rank space. We define the $\Delta$-ary segment wavelet tree as a recursive decomposition of the line segments in $\segments$ similar to the $\Delta$-ary wavelet tree. 
For simplicity, we assume that $n$ is a power of $\Delta$.
Define $\segments[a,b] \subseteq \segments$ to be the segments with $y$-coordinate in $[a,b]$. The \emph{$\Delta$-ary segment wavelet tree} $T$ for $
\segments$ is a complete balanced $\Delta=\lceil\lg^\epsilon{n}\rceil$ tree on $n$ leaves.
Each node $v$ represents a set of segments $\segments(v) = \segments[a,b]$.
The root $r$ represents $\segments(r) = \segments[1,n] = \segments$.
Let $v$ be an internal node representing the segments $\segments(v) = \segments[a,b]$, and let $v_0,\ldots, v_{\Delta-1}$ be the children of $v$.
Then $v_i$ represents the segments $\segments(v_i) = \segments[a + \lfloor(1+b-a)/\Delta \cdot i\rfloor, a + \lfloor(1+b-a)/\Delta \cdot (i+1)\rfloor - 1]$.
A leaf node $v$ represents a single segment $\segments(v) = \segments[a,a]$.
We store the $\Delta$-ary segment wavelet tree succinctly as follows. 
Intuitively, each internal node $v$ stores the structure of \cref{lemma:narrowgrid},
but to achieve the desired space complexity, we instead, for each level in the $\Delta$-ary segment wavelet tree, concatenate the segments of the nodes on that level into one single structure of \cref{lemma:narrowgrid}, as in \cite{DBLP:journals/tcs/MakinenN07}.
Since there is exactly one segment of each $y$ coordinate, we can easily maintain the indices into the single structure corresponding to each node.
When $n$ is not a power of $\Delta$, we pack the segments such that all levels in $T$ are complete except possibly the lowest, which is filled from the left.
We then store the path along with indices to the rightmost leaf on the lowest level.
This allows us to compute the indices into the single structure of each level.

\subparagraph*{Analysis} Each level of the tree contains all the segments and thus uses $2n\lg{\lg^\epsilon{n}} + O(n)$ bits of space by \cref{lemma:narrowgrid}.
Since the height of this tree is $\log_{\Delta}{n}=\lg{n}/\lg{\lg^\epsilon{n}}$ the entire data structure uses $2n\lg{n} + O(n\lg{n}/\lg{\lg{n}}) = 2n\lg{n} + O(n\lg{n}/\lg{\lg{n}}) $ bits of space and $O(n\lg{n}/\lg{\lg{n}})$ preprocessing time.

\subsection{Segment Access Queries}
We now show how to answer
$\segaccess$ queries in $O(\lg{n}/\lg{\lg{n}})$ time.
To answer a $\segaccess(y)$ query,
we first compute the path to the $y$th leaf in the left-to-right order.
We can trivially compute the path by performing a top-down traversal of the $\Delta$-ary segment wavelet tree $T$ starting at the root and ending in the $y$th leaf.
We then do a bottom-up traversal of the $\Delta$-ary segment wavelet tree $T$ starting at the $y$th leaf. 
Let $s = (x_l, x_r, y)$ denote the segment we are searching for. At each node $v$ in the traversal we maintain the following \emph{local variables}:
\begin{align*}
    l_v &= \text{the number of endpoints in $\segments(v)$ who has $x$-coordinate at most $x_l$.}\\
    r_v &= \text{the number of endpoints in $\segments(v)$ who has $x$-coordinate at most $x_r$.}
\end{align*}
We perform the traversal as follows.
Initially, $v$ is the $y$th leaf and $l_v = 1$ and $r_v = 2$.
Consider a non-root node $v$ with parent $v_p$ and let $v$ be the $k$th child of $v_p$.
We compute $l_{v_p} = \endsel(v_p, k, l_v)$ and $r_{v_p} = \endsel(v_p, k, r_v)$.
When we reach the root $u$ we return $(l_u, r_u, y)$.

\subparagraph*{Analysis} We use constant time at each node and visit one node at each level of the $\Delta$-ary segment wavelet tree. Since the height of $T$ is $O(\lg{n})$, the total time is $O(\lg{n}/\lg{\lg{n}})$.

\subparagraph*{Correctness}
Let $s=(x_l, x_r, y)$ be the segment with $y$-coordinate $y$.
We show inductively that $l_v$ and $r_v$ are computed correctly for each $v$ on the path in the bottom-up traversal.
When $v$ is the $y$th leaf we have $\segments(v)=\{s\}$ and thus $l_v=1$ and $r_v=2$.
Consider an internal non-root node $v$.
Assume $s\in\segments(v)$ and $l_v$ and $r_v$ are correct.
Let $v_p$ be the parent of $v$ and let $v$ be the $i$th child of $v_p$.
The segments in $\segments(v)$ are in slab $i$ of $v_p$.
Thus by the definition of $\endsel$ $l_{v_p} = \endsel(v_p, b, l_v)$ is the number of endpoints in $\segments(v_p)$ who has $x$-coordinate at most $x_l$ and $r_{v_p} = \endsel(v_p, b, r_v)$ is the number of endpoints in $\segments(v_p)$ who has $x$-coordinate at most $x_r$.
When we arrive at the root $u$ we have $x_l=l_u$ and $x_r=r_u$.

\subsection{Segment Select Queries} 
\label{sec:multisegment_wavelet:select}
We now show how to answer 
$\segsel$ queries in $O(\lg{n}/\lg{\lg{n}})$ time.
To answer a $\segsel(i,j)$ query, we do a top-down traversal in the $\Delta$-ary segment wavelet tree $T$  starting at the root and ending in the leaf containing the $j$th crossing segment of the vertical line at time $i$.

At each node $v$ with $\segments(v) = \segments[a,b]$ in the traversal we maintain the following \emph{local variables}:
\begin{align*}
    i_v =& \text{the number of endpoints in $\segments(v)$ who has $x$-coordinate at most $i$.}\\
    \bar{j}_v =& \text{the number of segments in $\segments[1, a-1]$ crossing the vertical line at $i$.}\\
    j_v =& \text{$j - \bar{j}_v$, such that the $j_v$th segment in $\segments(v)$ crossing the vertical line at $i$}\\
    &\text{is the $j$th segment in $\segments$ crossing the vertical line at $i$.} 
\end{align*}
We perform the traversal as follows. 
Initially, $v$ is the root and we have $i_{v}=i$, $\bar{j}_{v}=0$ and $j_{v}=j$.
Consider an internal node $v$ with children $v_0, \ldots, v_{\Delta - 1}$ and suppose we have computed $l_v$, $\bar{j}_{v}$, and $j_v$. 
We compute the child $v_k$ containing the $j_v$th segment crossing the vertical line at $i_v$ as
\[
    k=\slabsel(v, i_v, j_v)
\]
We then compute the local variables for the child $v_k$ as
\begin{align*}
    i_{v_k} &= \endrank(v, k, i_v)\\
    \bar{j}_{v_k} &= \slabrank(v, i_v, k-1) + \bar{j}_v\\
    j_{v_k} &= j - \bar{j}_{v_k}
\end{align*}
When $v$ is a leaf and $\segments(v)=\segments[a,a]$ we return $a$.

\subparagraph*{Analysis} We use constant time at each node and visit one node at each level of the $\Delta$-ary segment wavelet tree. Since the height of $T$ is $O(\lg{n}/\lg{\lg{n}})$, the total time is $O(\lg{n}/\lg{\lg{n}})$.

\subparagraph*{Correctness} 
Let $s$ be the $j$th smallest segment crossing the vertical line at $i$. 
We show inductively that $s\in \segments(v)$ and $i_v$, $\bar{j}_v$, and $j_v$ are computed correctly for each $v$ on the path in the top-down traversal.

When $v$ is the root we have $s\in \segments(v) = \segments$ and $\bar{j}_v=0$ and $j_v=j$. By definition, we have $i_v= i$ is the number of endpoints in $\segments$ with $x$-coordinate at most $i$.

Consider an internal non-root node $v$. Assume $s\in \segments(v)$ and that $i_v$, $\bar{j}_v$, and $j_v$ are correct. Let $v_k$ be the child of $v$ computed by the algorithm. 
The segments in $\segments(v_k)$ are in slab $k$ of $v$.
By the definition of $\slabsel$ the $j_v$th segment crossing the vertical line at $i_v$ is $k=\slabrank(v, i_v, k-1)$.
Furthermore by the definition $\slabrank(v, i_v, k-1)$ is the number of segments crossing the vertical line at $i$ in $\segments(v_0) \cup \ldots \cup \segments(v_{k-1})$ and $\endrank(v, k, i_v)$ is the number of endpoints in $\segments(v_k)$ who has $x$-coordinate at most $i$.
Hence $i_{v_k} =\endrank(v, k, i_v)$ and $\bar{j}_{v_k} = \slabrank(v, i_v, k-1) + \bar{j}_v$.

\subsection{Segment Rank Queries}
We now show how to answer $\segrank$ queries in $O(\lg{n}/\lg{n}\lg{n})$ time.
To answer a $\segrank(i, y)$, we perform a top-down traversal in the segment wavelet tree $T$ starting at the root and ending in the leaf $v$ such that $\segments(v)=\segments[y,y]$.
At each node $v$ with $\segments(v)=\segments[a,b]$ in the traversal, we compute the local variables $i_v$, and $\bar{j}_v$ in the same manner as in \cref{sec:multisegment_wavelet:select}.
We perform the traversal as follows.
Consider an internal node $v$ with $\segments(v)=\segments[a,b]$ and children $v_0, \ldots, v_{\Delta-1}$.
we continue the traversal in child $v_k$, where $k=\lfloor(y-a) / ((1+b-a) / \Delta)\rfloor$.
When $v$ is a leaf we return $\bar{j_v} + 1$ if $i_v=1$ and $\bar{j_v}$ otherwise.

\subparagraph*{Analysis} We use constant time at each node and visit one node at each level of the segment wavelet tree. Since the height of $T$ is $O(\lg{n}/\lg{\lg{n}})$, the total time is $O(\lg{n}/\lg{\lg{n}})$.

\subparagraph*{Correctness}
The correctness follows immediately from the correctness argument in \cref{sec:multisegment_wavelet:select} and the definition of $i_v$, $\bar{j_v}$ and $\slabrank$.

In summary, we achieve \cref{theorem:segment_select_fusion}.

%% file: 05-lower-bounds.tex
\section{Lower Bounds}\label{section:lowerbounds}
\subsection{Horizontal Line Segments in Rank Space}\label{section:worst-case-entropy}
Here, we show \cref{theorem:segment:succinct}.
Let $\segments$ be a set of $n$ horizontal line segments in rank space on the plane $[1,2n]\times [1,n]$ such that there is exactly one endpoint on each $x$-coordinate and one segment on each $y$-coordinate.
If we only consider the $x$-coordinates of the left and right endpoint of each segment, then the number of ways to arrange $n$ pairs of endpoints is the number of ways that we can pair $2n$ elements.
\begin{align*}
    \prod_{i=1}^{n}(2i-1) = \frac{(2n)!}{2^n n!}
\end{align*}
Since each segment has a unique $y$-coordinate,
the number of ways the segments can be arranged on the $y$-axis is $n!$.
Thus, the number of ways to arrange $n$ segments in rank space $[1,2n]\times[1,n]$ is $\frac{(2n)!}{2^n}$.
Thus to distinguish between each instance we need $\left\lceil\lg{\frac{(2n)!}{2^n}}\right\rceil=2n\lg{n} - O(n)$ bits.
In summary, we have shown \cref{theorem:segment:succinct}.

\subsection{Segment Rank in Rank Space}
Here, we show \cref{tm:segrank_lowerbound}.
We reduce from \emph{2-dimensional dominance counting}.
The 2-dimensional dominance counting problem is to preprocess $n$ points $(x_1, y_1), \ldots, (x_n, y_n)$ from rank space, such that given a point $(x,y)$ compute the number of points $(x_i,y_i)$ such that $x_i \leq x$ and $y_i\leq y$.
\begin{lemma}[\cite{DBLP:conf/stoc/Patrascu07}]
Any static data structure on $n$ points that uses $n\lg^{O(1)}{n}$ bits of space requires $\Omega(\frac{\lg{n}}{\lg{\lg{n}}})$ time to support dominance counting queries.
\end{lemma}
Given $n$ points $(x_1, y_1), \ldots, (x_n, y_n)$ to the 2-dimensional dominance counting problem, we construct an instance of the segment representation problem.
We assume wlog. that these $n$ points are in rank space on the plane $[1,n]\times[1,n]$ such that there is exactly one point on each $x$ and $y$ coordinate.
To see why this assumption is acceptable to establish the lower bound, we refer to the discussion by Pătrașcu~\cite{DBLP:conf/stoc/Patrascu07}.
We construct the $n$ segments $\segments$ as follows:
For each point $(x_i, y_i)$ we construct the corresponding segment such that the left endpoint is $(x_i, y_i)$ and the right endpoint is $(n+x_i, y_i)$ for $1\leq i \leq n$.
For a given query point $(x,y)$, we count the number of points $(x_i, y_i)$ such that $x_i\leq x$ and $y_i\leq y$ by performing the query $\segrank(x, y)$.
Since no segment ends before $x$-coordinate $n$, the number of segments crossing the vertical line through $x$-coordinate $x$ with $y$-coordinate in $[1,y]$ are the segments with left endpoints $(x_i, y_i)$ such that $x_i \leq x$ and $y_i \leq y$.
In summary, we have shown \cref{tm:segrank_lowerbound}.